\documentclass[a4paper,twoside]{article}
\usepackage{amsthm,amscd}

\usepackage{amsmath,amssymb,amsfonts}
\usepackage[numbers]{natbib}

\usepackage{graphicx}               
\usepackage{color}                  
\usepackage[absolute]{textpos} 
\usepackage{multicol}
\usepackage{array}

\theoremstyle{remark}

\usepackage[T1]{fontenc}

\usepackage{xcolor}

\newcommand{\beq} {\begin{eqnarray*}}
\newcommand{\eeq} {\end{eqnarray*}}

\def \E{\mathbb{E}}
\def \P{\mathbb{P}}

\DeclareMathOperator*{\argmin}{Argmin}

\newcommand{\X}{\textbf{X}}
\newcommand{\Y}{\textbf{Y}}

\newcommand{\uu}{\textbf{u}}
\theoremstyle{plain}

\newtheorem{ass}{Assumption}[section]

\newtheorem{rmk}{Remark}[section]
\newtheorem{defi}{Definition}[section]

\begin{document}

\title{New sensitivity analysis subordinated to a contrast}

\author{Jean-Claude Fort \thanks{Universit\'e Paris Descartes, SPC, MAP5, 45 rue des Saints P\`eres, 75006 Paris, France}, Thierry Klein\thanks{Institut de
Math\'ematiques de Toulouse, Universit\'e Toulouse 3, 31062 Toulouse C\'edex 9, France}, Nabil Rachdi\thanks{EADS Innovation Works, 12 rue Pasteur, 92152 Suresnes}}

\maketitle

\begin{abstract} In a model of the form $Y=h(X_1,\ldots,X_d)$ where the goal is to estimate a parameter of the probability distribution of $Y$, we define new sensitivity indices which quantify the importance of each variable $X_i$ with respect to this parameter of interest. The aim of this paper is to define {\it goal oriented sensitivity indices} and we will show that Sobol indices are sensitivity indices associated to a particular characteristic of the distribution $Y$. We name the framework we present as {\it Goal Oriented Sensitivity Analysis} (GOSA).

\end{abstract}
\ \\
{\bf Mathematics Subject Classification:} \\
{\bf Keywords:} Sensitivity analysis, Sobol indices 

\section*{Introduction}

From more than one decade, uncertainty propagation and sensitivity analysis are widely used to handle mathematical models of industrial problems involving many parameters or variables (e.g. see \cite{saltelli2004sensitivity}): geophysics and oil reservoir, safety in nuclear industry, soil pollution, and more generally domains where it can be found heavy computation codes with large number of inputs and complex computations so that only few simulations of these codes can be run (for agricultural example see \cite{Monod2006}).

The uncertainty propagation methodology (see \cite{de2008uncertainty} for more details) uses random variables as inputs, even for deterministic codes, and study the distribution (or some characteristics) of the output. It is justified on the one hand by the poor knowledge of the input parameter (or variables), and on the other hand by the relatively small number of observed output available. 

Very often, some of the input variables strongly affect the output (or a characteristic), while others have a small effect (and even no effect). The sensitivity analysis try to quantify these effects. In \cite{sobol} I.M. Sobol defined indices, now called Sobol indices, based on the decomposition of the output variance. Using the ANOVA decomposition of a function of several variables he defined global and partial indices for one or a group of variables.

Sensitivity Analysis is of great interest in industrial applications where the engineers deal with heavy computer codes, often with a large input dimension (about 50 in most of simulations). As a matter of fact, it is necessary to reduce the input dimension in order to make the simulations more tractable for the study of interest (thermal, acoustic, electromagnetic study...). Such sensitivity analysis may be done for instance thanks to experts judgement. In this case the experts choose the inputs to fix to a "nominal" value and consider the other ones as "free". A modelling phase is guided by some requirements which give some characteristics to be satisfied by the simulation. So, the design of a modelling (simulation code), in particular the choice of fixed inputs, may have to be done considering these requirements. As an example, let us consider a flight of a commercial aircraft where one of the requirements is that the pressure inside the cabin must be greater than a specific threshold (about 0.7 bar). The Environmental Control System (ECS) of an aircraft is a system providing air supply, thermal control and cabin pressurization for the crew and the passengers, which  guarantees their comfort. During the design of a new aircraft, the unique way to forecast the cabin environment is to simulate models representing the ECS, often very complex. Hence, an important challenge it to have a "good fidelity" of the ECS modelling so as to make pertinent simulations which will be at the heart of crucial decisions. The ECS modelling can be basically viewed as a black box input/output computer code with a large input dimension. Therefore, it is unavoidable to reduce this dimension by considering only "important" variables. As said before, a first approach may be to use experts judgement. Then, a very popular and widely used approach is to compute Sobol indices which give the contribution of the variability of each input w.r.t the outputs. At this point, one can wonder if the systematic use of Sobol index, whatever the quantity of interest, is the best thing that can be done. Indeed, it might be judicious to adopt a new approach based on the goals of the study. In particular, this new approach would be to consider the goal of this study example which is to verify if the cabin pressure is greater than 0.7 bar, and then to select the "important" variables w.r.t these requirements.
In this paper we will focus on this latter approach (which is different from the Sobol one) and we will propose a methodology to build new sensitivity indices based on the quantity of interest to estimate.\\
We aim at developing an approach we name {\it Goal Oriented Sensitivity Analysis}. This first study aims at proposing new global indices for one or several variable(s) which generalizes the Sobol ones. The general idea of this work is the following, the importance of an input variable may vary depending on what the quantity of interest is. \\

The paper is organized as follows.  Section \ref{section:discussion} presents a short discussion explaining why our new index is interesting. Then  we give a motivation of our work starting from the definition of Sobol indices in Section \ref{sec-motivation}. We define the notion of contrast function in Section \ref{contrast} and provide some classical contrasts.  In Section \ref{sensitivity} we define a new index which is contrast adapted.  Then in Section   \ref{ex}, we present two simple examples in order to illustrate our index, its properties and to show that he provides in some cases more information than the classical Sobol index. Section \ref{maxvrais}  is devoted to the case of the maximum likelihood estimation where we also present a simple analytical example as an illustration. Section \ref{est} is devoted to the practical estimation of our indices which is applied to two numerical examples one of them being the study of the classical "Ishigami" function.

\section{Sensitivity indices and goal oriented estimation}\label{section:discussion}

The Sobol indices have been widely used in many contexts. Application studies generally show a common drawback : they do  not emphasized a capital point, namely that the efficiency of an index has to be ranked  $w.r.t.$ the statistical parameter(s) or features that have to be estimated.

It seems very intuitive that to estimate  a mean or a median (central parameter) could involve very different variables than estimating  extreme quantiles. Thus the same index should not be used for these two different tasks.  So we need to adapt the indices to each particular goal we track, that we may call a "goal oriented" sensitivity study. As a matter of fact the Sobol indices are well suited to quantify the sensitivity of an estimator based on a variance criterion : a mean.

Shortly speaking, we propose to define an index for each statistical purpose.

Of course it may happen that several goals are to be reached, then one can adopt a mixed strategy i.e compute various indices related to each goal and combine them to define some importance criteria of the input variables. \\

\section{Motivations}\label{sec-motivation}
 
Let us first recall some well known facts about Sobol index. In a model $Y=h(X_1,\ldots,X_d)$ the global Sobol index quantify the influence of a random variable $X_i$ on 
the output $Y$. This index is based on the variance (see \cite{sobol},\cite{saltelli2000sensitivity}): more precisely, it compares the total 
variance of $Y$ to the expected variance of the variable $Y$ conditioned by $X_i$,
\begin{equation}\label{sobol2}
S_i=\frac{\text{Var}(\mathbb E[Y|X_i])}{\text{Var}(Y)}.
\end{equation}
By the property of the conditional expectation it writes also 
\begin{equation}\label{sobol1}
S_i=\frac{\text{Var}(Y)-\mathbb E(\text{Var}[Y|X_i])}{\text{Var} Y}.
\end{equation}
Formula (\ref{sobol2}) is generally used by people working in the domain of uncertainty analysis, see \cite{sobol,saltelli2002making,saltelli2000sensitivity,saltelli2004sensitivity}.

We propose to adopt formula (\ref{sobol1}) to extend the definition of a global Sobol index according to the estimation of a parameter.

Indeed, it is well known that the mean $\mathbb EY$ is the minimizer of the quadratic function $\theta \mapsto \mathbb E(Y-\theta)^2$ (we will call it a {\it contrast function} later) and that the value of the minimum is the variance of $Y$. Now, conditioning by $X_i$, $\mathbb E[Y|X_i]$ is the minimizer of the function $\mathbb E[(Y-\theta)^2|X_i]$ and the minimum value  is $\text{Var}[Y|X_i]$. So that $S_i$ appears to compare the optimal value of the function $\mathbb E(Y-\theta)^2$ to the expected optimal value of the conditional function $\mathbb E[(Y-\theta)^2|X_i]$.\\
This remark will guide our definition of a new index associated to a given contrast. Indeed, if one replace the contrast function $\theta \mapsto \mathbb E(Y-\theta)^2$ by an another contrast, he will obtained naturally an another index.  

\section{Notion of contrast function}\label{contrast}

\subsection{Definition}

{\defi Let $\Theta$ be some generic set and $Q$  be some probability measure on a space $\mathcal{Y}$. A $(\Theta, Q)$-{\bf contrast function}, or simply {\bf contrast function}, is defined as any
function $\psi$
\begin{eqnarray}
  \psi \, : \, \Theta &\longrightarrow& L_{1}(Q) \\
    \nonumber    \theta &\longmapsto& \psi(\cdot, \theta) \, : \, y\in\mathcal{Y} \longmapsto \Psi(\rho,y) \,, \label{def_contrast}
\end{eqnarray}
such that
\begin{eqnarray}
\theta^{*} = \argmin_{\theta\in\Theta} \mathbb{E}_{Y \sim Q}\,\psi(Y; \theta)\, \label{charac_feat}
\end{eqnarray}
 is {\it unique}. The function $\Psi \, : \, \theta \mapsto \mathbb{E}_{Y \sim Q}\,\psi(Y; \theta)$ is the {\bf average contrast function}, or abusively contrast function if there is no ambiguity.\\

}

The contrast function is a very useful object in {\it Statistical Learning Theory} (see \cite{massart2007concentration}) where it defines estimation procedures of some feature $\theta^{*} \in \Theta$ (scalar or functional) associated to a random variable $Y$. For instance, when observing a  $n$-sample $(Y_1,\ldots,Y_n$) of the random variable $Y$, an estimator of $\theta^{*}$ is given by $\widehat\theta=\mbox{Argmin}_{\theta} \Psi_n(\theta)$, where $\Psi_n$ is obtained by substituting the expectation $w.r.t.$ the variable $Y$ by the expectation $w.r.t.$ the empirical measure of the sample. It reads:
$$\Psi_n(\theta)=\frac{1}{n}\sum_{i=1}^n \psi(Y_i;\theta).$$
Then, considering for example the following contrast function
 $$ \psi \,:\, (y; \theta)\mapsto (y - \theta)^2 $$
provides the well known estimation procedure of the mean
$$ \widehat\theta=\mbox{Argmin}_{\theta} \frac{1}{n}\sum_{i=1}^n (Y_i - \theta)^2 \,. $$
That gives obviously
$$ \widehat\theta = \frac{1}{n}\sum_{i=1}^n Y_i \,.$$
The same stands for "functional" features associated to the random variable $Y$ (density function, etc.). 

{\rmk \label{rmk-char-cont} {\bf Feature characterization by contrast.} The writing (\ref{charac_feat}) provides a characterization of a feature $\theta^*$ of $Y$ by a contrast $\psi$. Notice that it may exist various contrasts $\psi$ characterizing $\theta^*$ (see \cite{Phdrachdi} for more details).\\

}

In this paper, we do not use contrast functions in order to estimate a feature of $Y$ but rather for defining new sensitivity indices as we will see. Indeed, our aim is to define sensitivity indices that rely on specific features of a random variable of interest $Y$: sensitivity w.r.t the mean, w.r.t an $\alpha$-quantile, w.r.t the density function, etc. For this, we will use the contrasts characterization and we will see that the Sobol indices are in fact particular indices associated to some particular contrast.

\subsection{Some examples of contrasts}

Let us give a non exhaustive list of contrasts that allow to estimate various parameters associated to a probability distribution. We give the classical contrasts associated to each parameter.

\begin{enumerate}
\item Central parameters: 
\begin{itemize}
\item The mean :  $\Psi(\theta)=\mathbb E |Y-\theta|^2$.
\item The median (in $\mathbb R$) : $\Psi(\theta)=\frac{1}{2}\mathbb E|Y-\theta|$.
\end{itemize}
\item An excess probability : $\Psi(\theta)=\mathbb E |{\bf1}_{Y\ge t}-\theta|^2$.
\item All the probability tail: $\Psi(\theta)=\int_{t_0}^\infty \mathbb E |{\bf1}_{Y\ge t}-\theta(t)|^2 dt$.
\item The $\alpha$-quantile : $\Psi(\theta)=\mathbb E (Y-\theta)(\alpha -{\bf1}_{Y\le\theta})$. 
\item All the quantile "tail": $\Psi(\theta)=\int_{\alpha_0}^1 \mathbb E (Y-\theta(\alpha))(\alpha -{\bf1}_{Y\le\theta(\alpha)})d\alpha$.
\item The probability density function, which is an infinite dimensional parameter.
\begin{itemize}
\item Using the kernel method with a given kernel $K$ and a window size $r>0$, we set $K_r(Y)=\frac{1}{r}K(\frac{Y}{r})$ and the contrast is:
$$\Psi(\theta)=\mathbb E \int_{-\infty}^{+\infty}(K_r(Y-t)-\theta(t))^2dt.$$
In fact  it is an unbiased estimator of the convolution of the $p.d.f$ with $K_r$, which is the target "parameter".
\item Using an orthonormal  $\mathbb L^2$ basis $(\varphi_j,j\ge 0)$ truncated at the order~$N$ :
$$\Psi(\theta)=\mathbb E \sum_{j=0}^N (\varphi_j(Y)-\int_{-\infty}^{+\infty} \varphi_j(u)\theta(u) du)^2.$$
What is really estimated here, is the orthogonal projection of the $p.d.f$ on the truncated basis, which is the target "parameter".
\end{itemize}
\end{enumerate}

Most of these contrasts are of quadratic type, contrary to the contrasts associated to the quantiles. 

\section{Sensitivity with respect to a contrast}\label{sensitivity} 

We are interested in the sensitivity of a scalar output $Y$ to an input variable $X_k$, we assume that $Y$ is a function of some input variables: $$Y=h(X_1,\ldots,X_d)=h({\bf X})\,. $$
Generally $h$ is a "black box", in the sense that $h$ is not explicit but results from heavy computer code, complex mathematical (or statistical) models.

{\rmk For sake of simplicity we consider a scalar output $Y$ but our method can easily be extended to a multiple output $Y\in \mathbb{R}^q$.\\ 
}

We assume that $\Psi$ is a contrast associated to a "parameter" 
 $\theta^*$, where $\theta^* =\mbox{Argmin} \Psi(\theta)$. Moreover $\Psi $ writes 
$\Psi(\theta)=\mathbb E\psi(Y;\theta)$.\\ 

We define the contrast variation with respect to the variable $X_k$.

\begin{defi}\label{variation}
Let $\Psi(\theta)=\mathbb E\psi(Y;\theta)$ be a contrast.
The contrast variation due to $X_k$ is defined as 
$$ V_k= \min_\theta \Psi(\theta)-\mathbb E( \min_\theta \mathbb E(\psi(Y;\theta)|X_k)),$$
and we have $V_k \geq 0$. We can also write $V_k$ as follows
\begin{equation}
V_k =  \mathbb{E}_{(X_k,Y)}\left(\psi(Y;\theta^*)  - \psi(Y;\theta_k(X_k))\right) \label{contrast_var}
\end{equation}
where $\theta^* = \displaystyle{\argmin_{\theta} \Psi(\theta) }$ and $\theta_k(x) = \displaystyle{\argmin_{\theta} \mathbb E(\psi(Y;\theta)|X_k = x)} \,. $
\end{defi}

Notice that the inequality $$\mathbb E(\min_\theta \mathbb E(\psi(Y;\theta)|X_k))\le \min_\theta \mathbb E(\mathbb E(\psi(Y;\theta)|X_k))= \Psi(\theta)$$
  implies that $V_k$ is non negative.\\  
Moreover, let us remark that if $Y$ does not depend on $X_k$, $V_k$ is $0$ and conversely if $Y=h(X_k)$ then $V_k$ is maximum.\\

We make the following assumption:
{\ass\label{positif} $$ \mathbb E \min_{\theta}\psi(Y;\theta)\in  \mathbb R.$$}
Notice that all the contrasts we presented in Section \ref{contrast} satisfy Assumption \ref{positif} since $\min_{\theta}\psi(Y;\theta)=0$.

Now we are in position to define a new index generalizing the Sobol one based on contrasts that satisfy Assumption \ref{positif}.   
  
\begin{defi} $\boldsymbol{\psi}${\bf-Indices.}\label{indice} Assume that a contrast $\Psi(\theta)=\mathbb E\psi(Y;\theta)$ satisfies Assumption \ref{positif}. The  $\psi$-index of the variable $Y=h(X_1,\ldots,X_d)$ with respect to the contrast $\Psi$ and the variable $X_k$ is defined as:
$$S^k_{\psi}=\frac{V_k}{ \min_\theta \Psi(\theta)-\mathbb E \min_{\theta}\psi(Y;\theta)},$$
or 
\begin{equation}
S^k_{\psi}= \frac{\mathbb{E}_{(X_k,Y)}\left(\psi(Y;\theta^*)  - \psi(Y;\theta_k(X_k))\right)}{ {\Psi(\theta^*)}-\mathbb E \min_{\theta}\psi(Y;\theta)}. \label{def_psi_indice}
\end{equation}
\end{defi}

\begin{bf}Let us make some comments on this new index\end{bf}
\begin{enumerate}
\item  If $Y$ does not depends on $X_k$, then $S^k_{\psi}=0$. Moreover, assuming that the variables $(X_1,\ldots,X_d)$ are independent, which is the case when people consider the Sobol indices,
if $Y=h(X_k)$ then $S^k_{\psi}=1$ and the other indices $S^l_{\psi},l\ne k$ are $0$. In fact, we have that 
$$ S^k_{\psi} \in [0,1] \,. $$

These basic properties were expected from a reasonable sensitivity index.

\item As mentioned in the introduction, when considering in (\ref{def_psi_indice}) the mean-contrast 
$$\psi\, :\, (y;\theta)\mapsto (y-\theta)^2\,, $$ 
we retrieve the global Sobol index. Indeed, in this case we have that 
\begin{eqnarray*}
\theta^* &=& \displaystyle{\argmin_{\theta} \Psi(\theta) }
=  \displaystyle{\argmin_{\theta} \mathbb{E}(Y-\theta)^2 }
= \mathbb{E}Y
\end{eqnarray*}
and
\begin{eqnarray*}
\theta_k(x) &=& \displaystyle{\argmin_{\theta} \mathbb E(\psi(Y;\theta)|X_k = x)}\\
&=&  \displaystyle{\argmin_{\theta} \mathbb E(Y - \theta)^2|X_k = x)}\\
&=& \mathbb{E}(Y|X_k=x) \,.
\end{eqnarray*}
Thus, it yields
\begin{eqnarray*}
\mathbb{E}_{(X_k,Y)}\left(\psi(Y;\theta^*)  - \psi(Y;\theta_k(X_k))\right) &=& \mathbb{E}_{(X_k,Y)}\left( (Y - \mathbb{E}Y )^2  - (Y - \mathbb{E}(Y|X_k))^2\right) \\
&=& \mathbb{E} (Y - \mathbb{E}Y )^2  - \mathbb{E}_{(X_k,Y)}(Y - \mathbb{E}(Y|X_k))^2\\
&=& \text{Var}(Y) - \mathbb{E}_{X_k}\mathbb{E}\left[(Y - \mathbb{E}(Y|X_k))^2 | X_k \right]\\
&=& \text{Var}(Y) - \mathbb{E}_{X_k} \text{Var}(Y|X_k)\\
&=& \text{Var}(\mathbb{E}(Y|X_k))
\end{eqnarray*}
and 
\begin{eqnarray*}
\min_\theta \Psi(\theta)-\mathbb E \min_{\theta}\psi(Y;\theta) &=& \Psi(\theta^*) - 0
= \text{Var}(Y)\,.
\end{eqnarray*}
Finally, we obtain the following $\psi$-index
$$ S^k_{\psi} = \frac{\text{Var}(\mathbb{E}(Y|X_k))}{\text{Var}(Y)} \,,$$
which is exactly the first order Sobol index defined in equation (\ref{sobol2}).
\end{enumerate}

{\rmk \label{rmk-gen_indice} The indices (\ref{def_psi_indice}) we propose can be generalized to higher order indices, for $I\subset\{1,...,d\}$ and denoting by $\X_I  =(X_i)_{i\in I}$, we define

\begin{equation}
S^I_{\psi} = \frac{\mathbb{E}_{(\X_I,Y)}\left(\psi(Y;\theta^*) - \psi(Y;\theta_I(\X_I))\right)}{\mathbb{E}_{Y}\psi(Y;\theta^*)} \label{ind_gen}
\end{equation}
where $\theta_I(\X_I)$ is the feature associated to the contrast $\psi$ of the random variable $Y|\X_{I}$.\\ 
}

In the next section we give two simple examples that allow to familiarize with our indices, and which highlight the differences between our indices and the Sobol ones.

\section{First examples}\label{ex} 

In this section, we present  two very simple examples in order to show 
\begin{enumerate}
\item how our indices can be analytically computed in very simple cases.
\item how they differ from classical Sobol indices. \\ Indeed in Example 1 (see Figure \ref{laplace}) one can see that Sobol indices are both constant and equal to .5 whereas the new indices are distinct and non-constant. One can note that Sobol indices give the same importance for each variable, and  the importance of each variable differs while $\alpha $ moves for the new indices (indeed for small values of $\alpha$ the second variable is more important than the first one and for large values of   $\alpha$ the first variable is more important than the second one). This example shows clearly that the new index is more adapted to the problem than the classical Sobol index. It seems very intuitive that for large values of $\alpha $ the positive variable has to be more important than the negative one.
\end{enumerate}
\subsection{Example 1} \label{sec-ex1}

\begin{figure}[!htbp]
\centering
\includegraphics[scale=0.45]{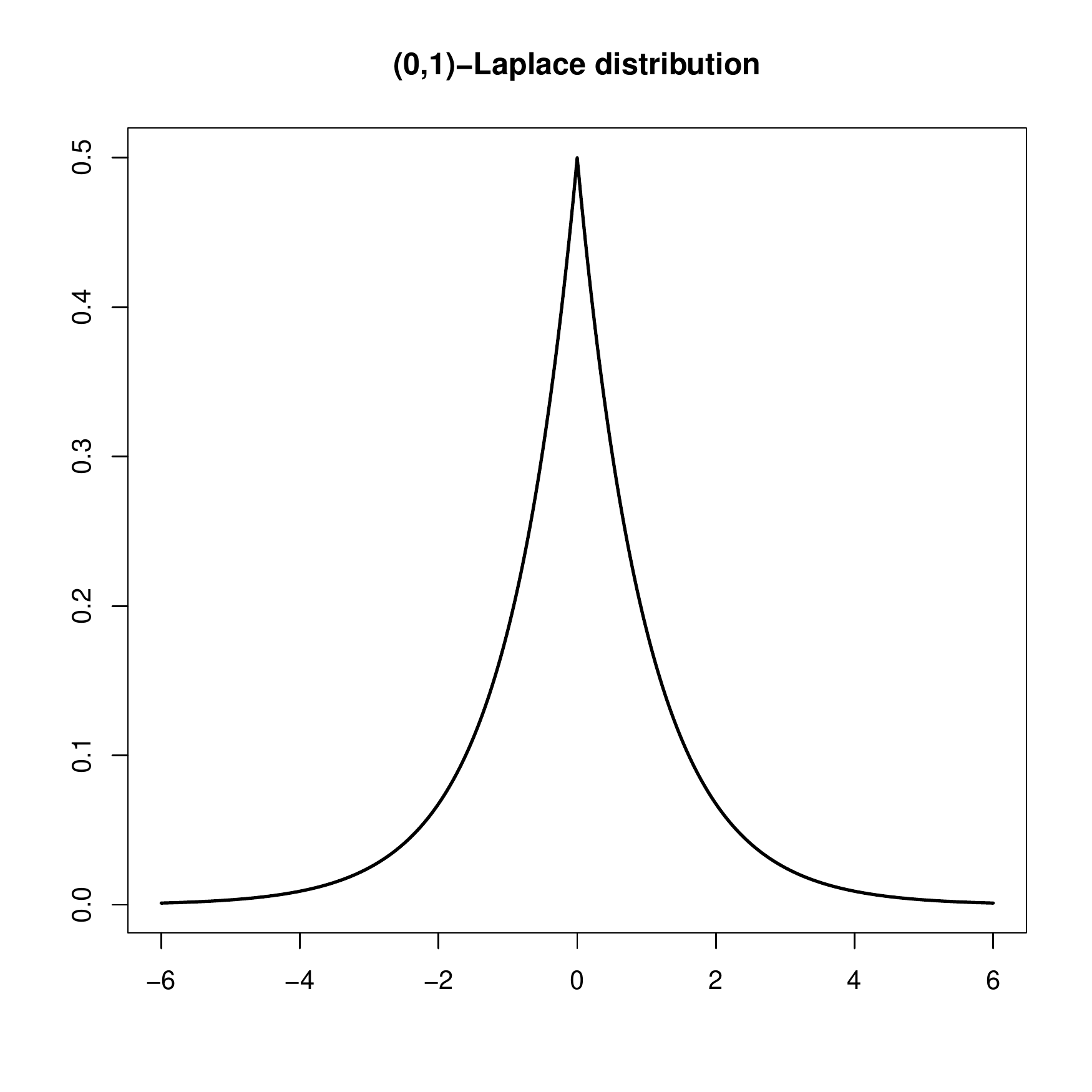}
\caption{Plot of (0,1)-Laplace distribution}
\label{fig:dist_laplace}
\end{figure}

Let $Y=X_1+X_2$, with $X_1\sim Exp(1), X_2\sim -X_1$, these two variables being independent. The variable $Y$ has the Laplace distribution  with parameter~$1$ its density with respect to the Lebesgue's measure is $f(x)=\exp(-|x|)/2$ (See Figure \ref{fig:dist_laplace}).\\ 
Here, we aim at defining sensitivity indices with respect to the $\alpha$-quantile $q_Y(\alpha)$ of $Y$. Thus, we propose to use the following contrast  
$$\Psi(\theta)=\mathbb E (Y-\theta)(\alpha -{\bf1}_{Y\le\theta})\,,$$
which characterises the $\alpha$-quantile.

Notice that with the previous contrast, $\mathbb E \min_{\theta}\psi(Y;\theta)=0$. Now, let us compute the indices
\begin{eqnarray*}
S^k_{\psi} &=& \frac{\mathbb{E}_{(X_k,Y)}\left(\psi(Y;\theta^*)  - \psi(Y;\theta_k(X_k))\right)}{\mathbb{E}\psi(Y;\theta^*)} \\
&=& \frac{\mathbb{E}\psi(Y;\theta^*)  - \mathbb{E}_{(X_k,Y)}\left(\psi(Y;\theta_k(X_k))\right)}{ \mathbb{E}\psi(Y;\theta^*)}\,, 
\end{eqnarray*}
for $k=1,\,2$, where $\theta^*= q_Y(\alpha)$, $\theta_1(X_1) = q_{Y/X_1}(\alpha) $ and $\theta_2(X_2) = q_{Y/X_2}(\alpha) \,.$ \\

A simple computation yields 
$$q_Y(\alpha)=\left\{\begin{matrix}
-\log(1-\alpha)-\log2 & if& \alpha\geq1/2,\\
\log(2\alpha)& if& \alpha<1/2.
\end{matrix}
\right.$$
and
$$\Psi(\theta)=\mathbb{E}\psi(Y;\theta)=\left\{\begin{matrix}
\frac{ \exp(\theta)}2-\theta\alpha& if& \theta<0,\\
\frac{ \exp(-\theta)}2-\theta\alpha+\theta& if& \theta\geq0.
\end{matrix}
\right.$$

We finally get the following indices (for detailed computations see Annexe \ref{annexe-ex1})
\begin{align}
S^1_{\psi}&=\left\{
\begin{matrix}\frac{(1-\alpha)(1-\log(2(1-\alpha)))+\alpha\log(\alpha)}{(1-\alpha)(1-\log(2(1-\alpha)))}&\mathrm{\ if\ }&\alpha\geq1/2\\
\frac{\alpha(1-\log(2\alpha))+\alpha\log(\alpha)}{\alpha(1-\log(2\alpha))}&\mathrm{\ if\ }&\alpha<1/2\\
\end{matrix}\right. \label{quant1_ex1}\\
S^2_{\psi}&=\left\{
\begin{matrix}
\frac{(1-\alpha)(1-\log(2(1-\alpha)))+(1-\alpha)\log(1-\alpha)}{(1-\alpha)(1-\log(2(1-\alpha)))}&\mathrm{\ if\ }&\alpha\geq1/2\\
\frac{\alpha(1-\log(2\alpha))+(1-\alpha)\log(1-\alpha)}{\alpha(1-\log(2\alpha))}&\mathrm{\ if\ }&\alpha<1/2
\end{matrix}\right. \label{quant2_ex1}
\end{align}

\begin{figure}[htbp!]
  \centering
 \fbox{
  \includegraphics[width=10cm]{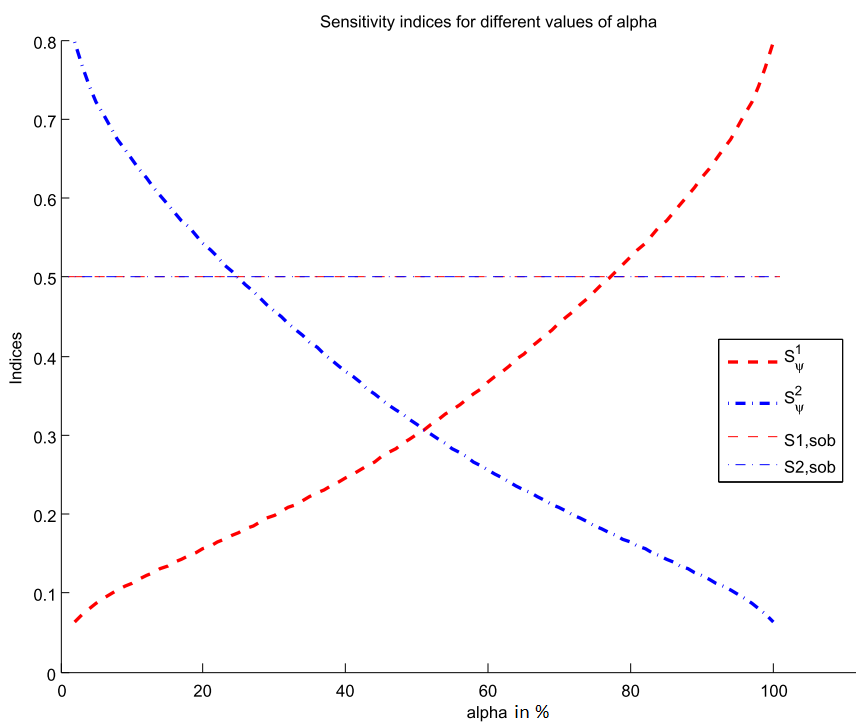}}\\
  \caption{Sensitivity Indices for Example 1.}\label{laplace}
\end{figure}

In Figure \ref{laplace} we plot the indices $S^1_{\psi}$ and $S^2_{\psi}$ and also the Sobol indices.

As it was expected,  we have $S^1_{\psi}<S^2_{\psi}$ for  $\alpha<1/2$,  $S^1_{\psi}>S^2_{\psi}$ for  $\alpha>1/2$ we and $S^1_{\psi}=S^2_{\psi}$ for $\alpha=1/2$. Moreover $\lim_{\alpha\to 1} S^1_{\psi}=\lim_{\alpha\to 0} S^2_{\psi}=1$ and  $\lim_{\alpha\to 0} S^1_{\psi}=\lim_{\alpha\to 1} S^2_{\psi}=0$.\\
In this example the corresponding classical Sobol indices are $S_{Sob}^1=S_{Sob}^2=1/2$. The latter indices do not depend on $\alpha$, in other words, the Sobol indices do not include the fact that we are investigating quantiles and hence they are useless if the statistical aim is to compute quantiles.

\subsection{Example 2}\label{section:laplace}

The second example is $Y=X_1+X_2$, with $X_1\sim Exp(1), X_2\sim Exp(a), a>0$, two independent variables. 
Here, we aim at providing a sensitivity index with respect to the probability of $Y$ to exceed $t\ge 0$, i.e $\mathbb{P}(Y\geq t)$. A contrast which characterises such quantity of interest can be the following 

\begin{equation}
\Psi(\theta)=\mathbb E |{\bf1}_{Y\ge t}-\theta|^2 \label{contrast_ex2}
\end{equation}

which in fact turns to be a quadratic contrast. Hence, we will recover the sobol index associated to the variable $Z={\bf1}_{Y\ge t}$. This fact deserves to be observed.\\
The density of $Y$ is 
for $a\neq1$:
$$
f(x)=\left\{ 
\begin{matrix}
 \frac{a}{a-1}\left(e^{-x}-e^{-ax}\right) \quad  &\text{for}& \, x\geq 0\\
 0 \quad &\text{else}& 
\end{matrix}
\right.
$$
and for $a=1$
$$ f(x) = xe^{-x} \quad  \text{for} \, x\geq 0 \,. $$

 \begin{figure}[!htbp]
\centering
\includegraphics[scale=0.45]{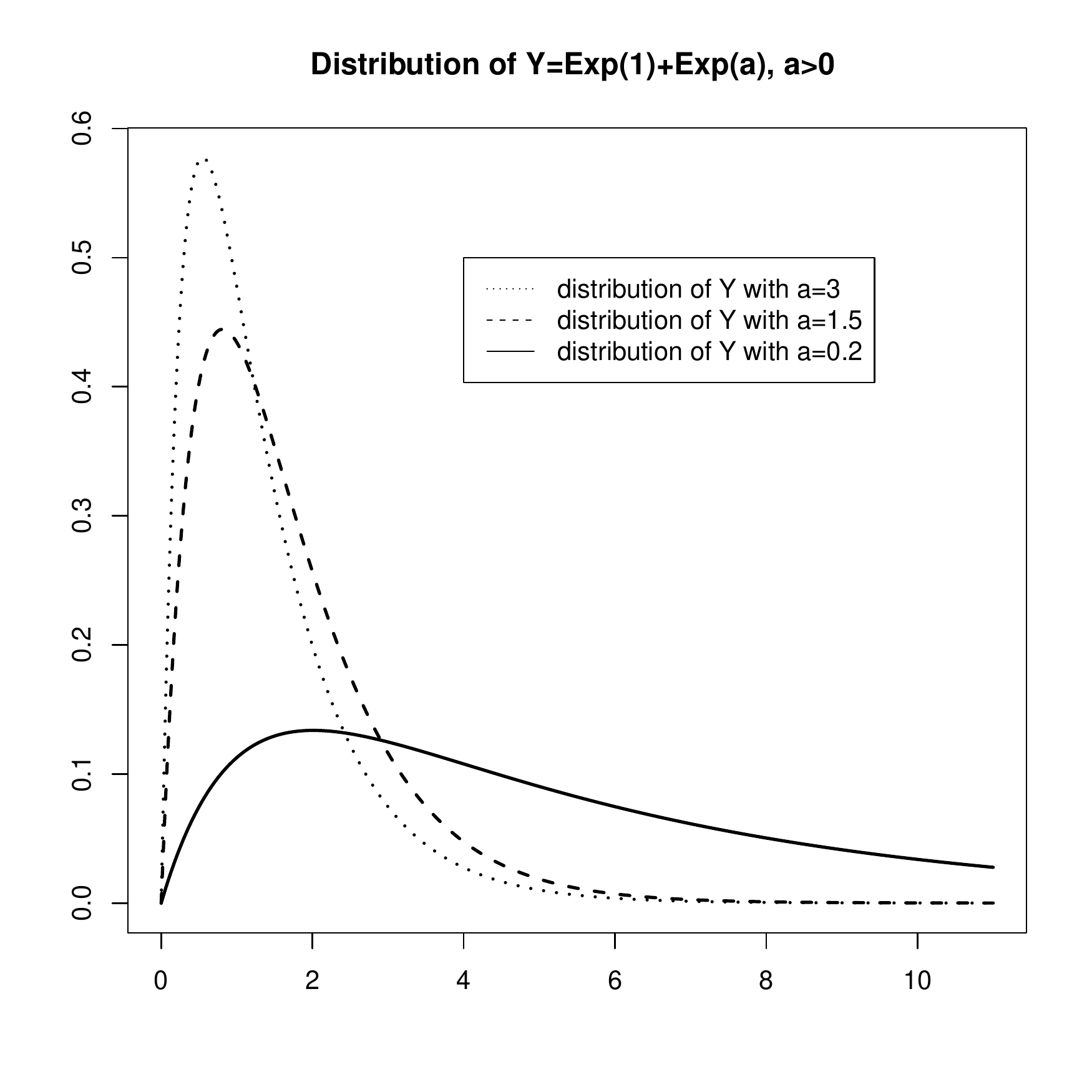}
\caption{Plot of the distribution of $Y$ for three values of $a$}
\end{figure} 

Notice that with the contrast (\ref{contrast_ex2}), one has $\mathbb E \min_{\theta}\psi(Y;\theta)=0$. Now, let us compute the indices
\begin{eqnarray*}
S^k_{\psi} &=& \frac{\mathbb{E}\psi(Y;\theta^*)  - \mathbb{E}_{(X_k,Y)}\left(\psi(Y;\theta_k(X_k))\right)}{ \mathbb{E}\psi(Y;\theta^*)}\,, 
\end{eqnarray*}
for $k=1,\,2$, where $\theta^*=\mathbb{P}(Y \geq t)$, $\theta_1(X_1) = \mathbb{P}(X_2 \geq t - X_1 / X_1 ) $ and $\theta_2(X_2) = \mathbb{P}(X_1 \geq t - X_2 / X_2 )\,.$ \\
First, we have easily
\begin{eqnarray*}
 \mathbb{E}\psi(Y;\theta^*) &=& \mathbb{E} ( {\bf1}_{Y\ge t} - \mathbb{P}(Y \geq t))^2\\
 &=& \text{Var}({\bf1}_{Y\ge t})\\
 &=& \mathbb{P}(Y \geq t)(1-\mathbb{P}(Y \geq t))\,.
\end{eqnarray*}
The computation of the quantities $E_k := \mathbb{E}_{(X_k,Y)}\left(\psi(Y;\theta_k(X_k))\right)$, $k=1,\,2$, is not as direct as the previous example. We compute it in Annexe \ref{annexe-ex2} and we obtain the following results
$$
E_1 =\left\{\begin{matrix}\frac{e^{-at}-e^{-t}}{1-a}-\frac{e^{-2at}-e^{-t}}{1-2a}&if &a\notin\{1,1/2\}\\
 e^{-2t}-(1-t)e^{-t}&if &a=1 \\
  2e^{-t/2}-(2+t)e^{-t}&if &a=1/2
\end{matrix}
\right.
$$

and
$$
E_2 =\left\{\begin{matrix}a\frac{e^{-at}-e^{-t}}{1-a}-a\frac{e^{-at}-e^{-2t}}{2-a}&if &a\notin\{1,2\}\\
e^{-2t}-(1-t)e^{-t}&if &a=1 \\
2e^{-t}- 2(t+1)e^{-2t}&if &a=2 \, .
\end{matrix}
\right.
$$

It is now easy to compute $S^1_{\psi}$ and $S^2_{\psi}$.\\
Next, considering the writing $Y = \xi_1 + \frac{1}{a}\xi_2$ where $\xi_1$ and $\xi_2$ are two independent variables with $\xi_1\sim Exp(1), \xi_2\sim Exp(1)$, it is easy to see that the Sobol indices are given by
 $$ S_{Sob}^1 = \frac{a^2}{1+a^2}\,, \quad S_{Sob}^2 = \frac{1}{1+a^2} \,.$$

\begin{figure}[htbp!]
  \centering
\fbox{
  \includegraphics[width=12cm, height=13cm]{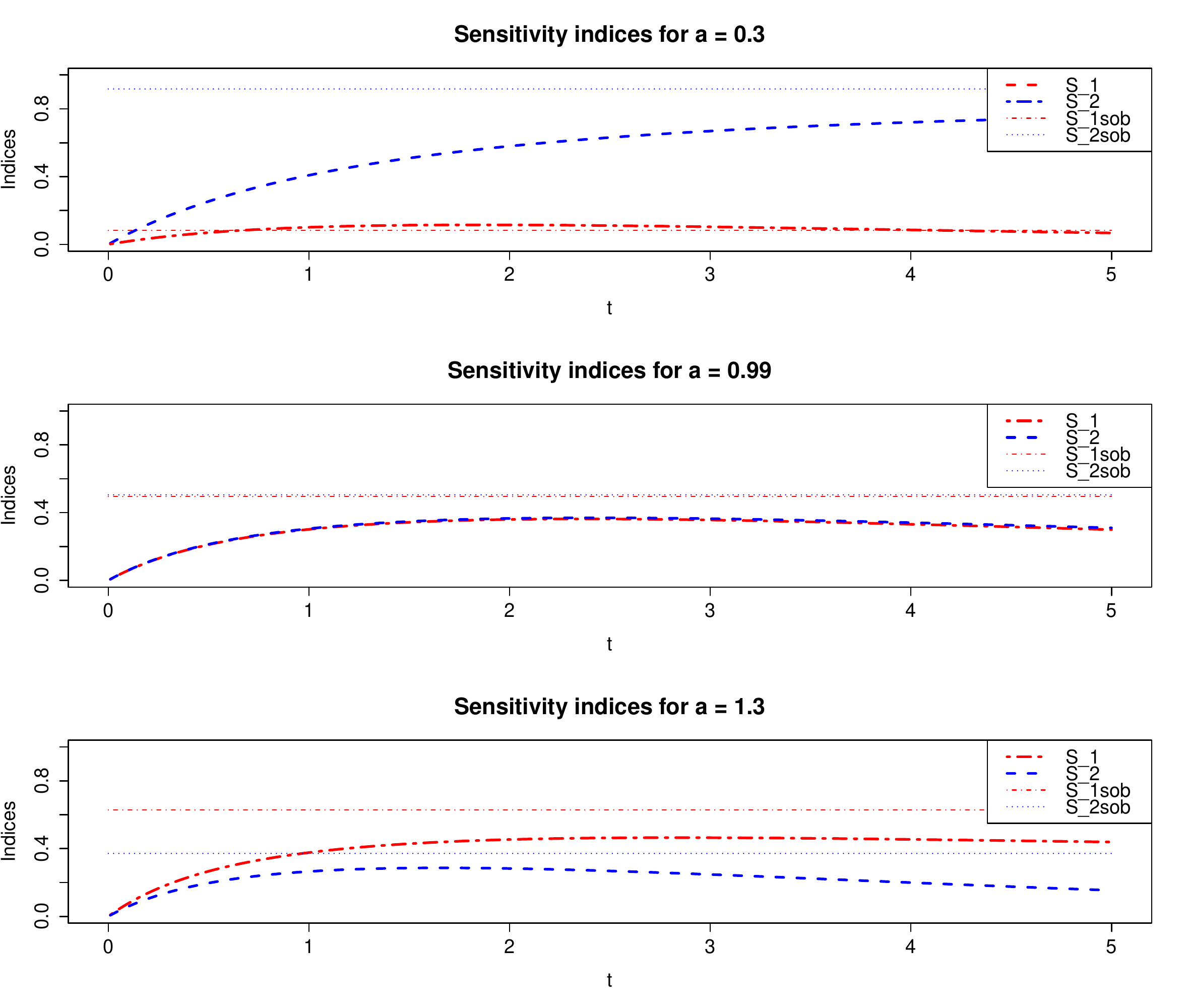}}\\
  \caption{Sensitivity Indices for Example 2 with $t\in [0,5]$, $a=0.3$, $a=0.99$ and $a=1.3$.}\label{laplace_a}
\end{figure}

We provide in Figure \ref{laplace_a} the plot of these indices in function of $t\in [0,5]$ for three values of $a$. 

These plots show first that the relative importance of each variable varies w.r.t to $t$ when considering the new indices, contrary to the Sobol ones. But, as it was mentioned at the beginning of this example, the contrast $\Psi(\theta)=\mathbb E |{\bf1}_{Y\ge t}-\theta|^2 $ induces in fact Sobol indices for the variable $Z={\bf1}_{Y\ge t}$, hence the importance ranking is preserved using the two types of indices. It was not the case for Example 1 where, for the new indices, the importance ranking changes with the quantile level, contrary to the Sobol ones.

\section{The case of the maximum likelihood}\label{maxvrais}
The use of maximum likelihood techniques in order to estimate a parameter is one of the most popular technique in statistics. That's why, in this section we present the classical example of maximum likelihood for which the contrast is not necessarily quadratic.

\subsection{A sensitivity analysis related to a maximum likelihood estimation }\label{MV} 

Let ${\bf X}$ be some random vector and assume that its distribution is known. Then, let us suppose that $Y$ is the result of a parametrized model $h({\bf X},\theta^*)$ that we can analytically handle. We want to estimate $\theta^*$. \\

For this, we use the maximum likelihood ($M.L.$) method: the estimation is based on the contrast $\Psi(\theta)=-\mathbb E \left(\log p_\theta(Y)\right)$ where for each $\theta$, $p_\theta(y)$ is the $p.d.f.$ of $Y$. Here, the expectation is taken with respect to the true distribution of $Y$. We  are now in position to define  a sensitivity index of the estimator of $\theta^*$.\\

Applying Definition \ref{variation} the variation of contrast $V_i$ is:
$$ V_i=\min_{\theta}\mathbb E\Big(-\log p_\theta(Y)\Big)-\mathbb E\Big(\min_{\theta}\mathbb E\big(-\log p_\theta(Y)|X_i\big)\Big).$$
Assuming that the $M.L.$ contrast satisfies Assumption \ref{positif}, we define:
\begin{equation}
S_i=\frac{\min_{\theta}\mathbb E\Big(-\log p_\theta(Y)\Big)-\mathbb E\Big(\min_{\theta}\mathbb E\big(-\log p_\theta(Y)|X_i\big)\Big)}
{\min_{\theta}\mathbb E\Big(-\log p_\theta(Y)\Big)-\mathbb E\Big(\min_{\theta}\big(-\log p_\theta(Y)\big)\Big)}
\label{ind_max_vrais}
\end{equation}

{\rmk The index $S_i$ may also writes
$$ S_i = \frac{\mathbb{E}_{(X_i,Y)}\left(\psi(Y;\theta^*)  - \psi(Y;\theta_k(X_i))\right)}{  \Psi(\theta^*)-\mathbb E \min_{\theta}\psi(Y;\theta)} \,,$$
where $\psi(y;\theta) = -\log( p_\theta)(y)\,.$ This latter expression is very useful when one knows the quantity of interest $\theta^*$, see Remark \ref{rmk-char-cont}. Here, or more generally for any model $Y=h(\X,\theta^*)$, the parameter $\theta^*$ is not directly defined as a specific feature of the distribution of $Y$ (like a quantile, the mean, etc.), and so it is not directly computable. Hence, in this case we prefer the writing (\ref{ind_max_vrais}).
}

\subsection{An analytical example}

As an analytical illustration we take the case of an exact linear gaussian model. Let $Y=\theta X_1+X_2$, where $\X=(X_1,X_2)$ has a normal distribution ${\cal N}(0, I_2)$. The distribution of $Y$ is ${\cal N}(0,\theta^2+1)$ and the maximum likelihood contrast reads:
$$\Psi(\theta)=\mathbb E \left(\frac{ Y ^2}{2(\theta^2+1)}+\frac{1}{2}\log( {2\pi}(\theta^2+1))\right).$$

The minimum is obtained for $\theta$ such that $\displaystyle\mathbb E Y ^2=\theta^2+1$ and has value 
$$\displaystyle \frac{1}{2}+\frac{1}{2}\log  {2\pi}+\log\mathbb E Y^2\,.$$ 

Likewise, conditionally to $X_i$ we have 
$$\displaystyle \mathbb E(-\log p_\theta(Y)|X_i)= \frac{\mathbb E\mathbb (Y ^2|X_i)}{\theta^2+1}+\frac{1}{2}\log( {2\pi}(\theta^2+1)).$$

The minimum is $\frac{1}{2}+\frac{1}{2}\log  {2\pi}+\log\mathbb E (Y^2|X_i)$.

 Now the variation $V_i$ is given by
 $$\displaystyle V_i=\frac{1}{2}(\log\mathbb E Y^2- \mathbb{E}(\log\mathbb E(Y^2|X_i)) ).$$

And finally we may write: 
$$S_i=\frac{\log  \mathbb EY^2-\mathbb E \left(\log \mathbb E( Y^2 |X_i)\right )}{\log  \mathbb E Y^2-\mathbb E \log Y^2}=\frac{\mathbb E\log(\frac{ \mathbb E ( Y^2 |X_i)}{\mathbb EY^2})}{\mathbb E\log(\frac{ Y^2}{\mathbb EY^2})}\,.$$
Thus we obtain the indices :

$$S_1=\frac{\mathbb E\log(\frac{1+\theta^2X_1^2}{1+\theta^2})}{\mathbb E\log\frac{Y^2}{1+\theta^2}}, \quad S_2=\frac{\mathbb E\log(\frac{\theta^2+X_2^2}{1+\theta^2})}{\mathbb E \log\frac{Y^2}{1+\theta^2}}\,$$
which we rewrite

$$ S_1=  \frac{1}{\gamma + \ln 2}\, \left( \ln (1+\theta^2) - \mathbb E\ln(1+\theta^2 \xi) \right), \quad S_2=  \frac{1}{\gamma + \ln 2}\, \left( \ln (1+\theta^2) - \mathbb E\ln(\theta^{2}+ \xi) \right)\, ,$$
where $\gamma$ is the Euler constant and $\xi$ is a chi-squared random variable with 1 degree of freedom Indeed, we have that $\frac{Y^2}{1+\theta^2}$ is a chi-squared random variable with 1 degree of freedom and one can verify that the expectation of its logarithm equals to $-(\gamma + \ln 2)$.\\

Moreover, one computes easily the Sobol indices

$$ S_{Sob}^1 = \frac{\theta^2}{1+\theta^2}, \quad S_{Sob}^2 = \frac{1}{1+\theta^2}\,.  $$

\begin{figure}[htbp!]
  \centering
\fbox{
  \includegraphics[scale=.5]{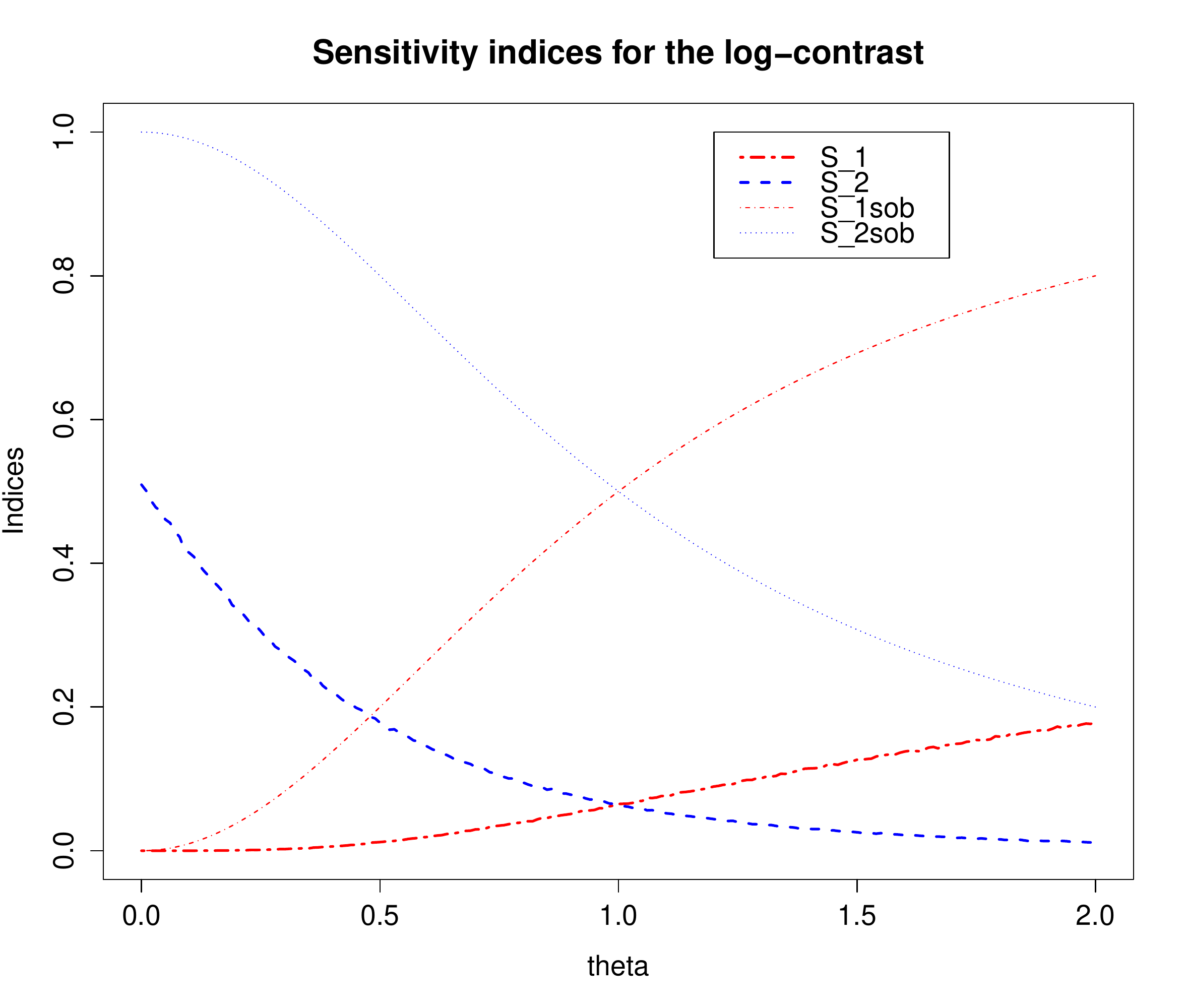}}\\
  \caption{Sensitivity Indices for the $\log$ contrast with $\theta\in [0,2]$.}\label{log_ind}
\end{figure}

\begin{figure}[ratio]
  \centering
\fbox{
  \includegraphics[scale=.5]{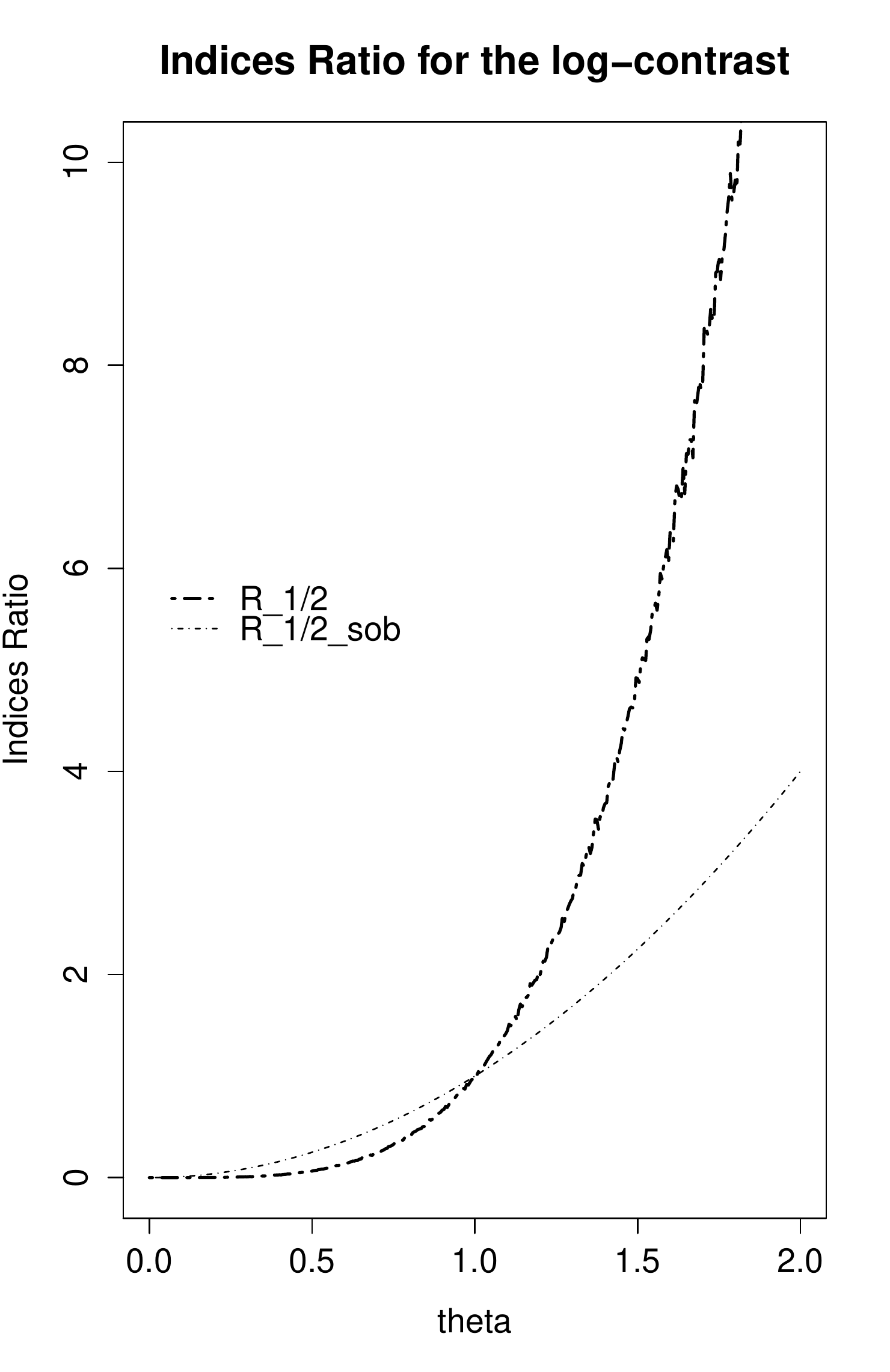}}\\
  \caption{ratios $\frac{S_1}{S_2}$ for the $\log$ contrast with $\theta\in [0,2]$.}\label{log_ind_ratio}
\end{figure}

Figure \ref{log_ind} shows that the contrast-based indices and the Sobol ones have the similar behaviour in that the importance ranking between $X_1$ and $X_2$ is exactly the same, with an equally importance for $\theta = 1$. The same conclusion was done for Example 2 where the contrast was "quadratic". Here, the contrast is not quadratic, but, for the Gaussian linear model considered, it acts like the mean contrast giving Sobol indices. Yet the ratios are not always of the same magnitude as can be seen in figure  \ref{log_ind_ratio}.  When $\theta$ is less than $1$ the ratios are closed, while for large $\theta$, $X_1$ appears to be more important for the  indices based on $\log$ contrast  than for the Sobol indices, which is almost intuitive.

\section{Estimation of the indices.}\label{est}
As in most application, the model $h$ is unknown or too complicated, it is useless to think that one can reach analytical formulas for our new indices. Hence, it is natural to wonder how these indices could be estimated. In this section, we first describe an estimation procedure and then we apply it for two examples. In order to validate our estimation procedure, we start by considering again Example 1 of Section \ref{sec-ex1}, and show that we indeed recover the theoretical values of these indices. Then we consider the case of the so-called Ishigami function (see \cite{ishigami1990importance}).

In this section, we present an estimation method for computing the contrast-based indices ($\psi$-indices)
\begin{equation*}
S^k_{\psi} = \frac{\Psi(\theta^*) - \mathbb{E}_{(X_k,Y)}\left(\psi(Y;\theta_k(X_k))\right)}{\Psi(\theta^*)}
\end{equation*}
or rather
\begin{equation}
S^k_{\psi} = \frac{\mathbb{E}_{(X_k,Y)}\left(\psi(Y;\theta^*) - \psi(Y;\theta_k(X_k))\right)}{\mathbb{E}_{Y}\psi(Y;\theta^*)} \label{ind_num}
\end{equation}
for some contrast $\psi$ (see Section \ref{contrast} for examples of contrasts), $\theta^* = \displaystyle{\argmin_{\theta} \mathbb{E}\psi(Y;\theta)}$ and $\theta_k(x) = \displaystyle{\argmin_{\theta} \mathbb E(\psi(Y;\theta)|X_k=x)} \,. $ Notice that here we considered contrasts that satisfied $\mathbb{E}\min_{\theta}\psi(Y;\theta)=0$, which is the case for our numerical applications.\\
By definition of the contrast function $\psi$, the parameter $\theta^*$ is a feature of the random variable $Y$, for instance:\\

\begin{itemize}
\item[$\bullet$] for the mean-contrast $\Psi(\theta)=\mathbb E |Y-\theta|^2$, $\theta^*$ is the expectation of $Y$, that is $\theta^* = \mathbb{E}(Y)$, and hence $\Psi(\theta^*) = \text{Var}(Y)$,
\medskip
\item[$\bullet$] for the $\alpha$-quantile contrast  $\Psi(\theta)=\mathbb E (Y-\theta)(\alpha -{\bf1}_{Y\le\theta})$, $\theta^*$ is the $\alpha$-quantile of $Y$, that is $\theta^* = q_{\alpha}(Y)$, and thus $\Psi(\theta^*)=\mathbb E (Y- q_{\alpha}(Y))(\alpha -{\bf1}_{Y\le q_{\alpha}(Y)})$,
\medskip
\item[$\bullet$] etc.\\
\end{itemize}

In fact, for numerical computations, $\theta^*$ may not be seen as an optimisation problem solution ($\theta^* = \displaystyle{\argmin_{\theta} \mathbb{E}\psi(Y;\theta^*)}$) but rather as the quantity of interest defined by the contrast $\psi$, see Remark \ref{rmk-char-cont}. Like this, we may have a direct computation of $\theta^*$ in many cases: mean, quantile, etc.\\
The same thing stands for the conditional parameter $\theta_k(x)$ which is a feature (depending on the considered contrast) of the random variable $Y|X_k=x$ where $X_k$ is fixed to $x$.\\

The computation of an estimator of the index (\ref{ind_num}) requires the two following steps (recall that $Y = h(X_1,...,X_p)$):\\

\begin{itemize}
\item [1] Generate $X^j_1,...,X^j_p$ and compute the $Y^j = h(X^j_1,...,X^j_p)$, for $j=1,...,n_1$. Then compute $\widehat{\theta^*}$ and replace in (\ref{ind_num}) the expectations $\mathbb{E}_{(X_k,Y)}$ and $\mathbb{E}_{Y}$ by their empirical versions.  

\item [2] Generate $X^{\prime\,j}_1,...,X^{\prime\,j}_p$ for $j=1,...,n_2$ (independent from the previous set) and compute the $Y^{\prime\,j} = h(X^{\prime\,j}_1,...,X^{\prime\,j}_p)$. Then, from the sample $Y^{\prime\,j}_k(x)=h(X^{\prime\,j}_1,...,X^{\prime\,j}_{k-1},x,X^{\prime\,j}_{k+1},...,X^{\prime\,j}_p)$, $j=1,...,n_2$, compute the function $x \mapsto \widehat{\theta}_k(x)$.\\
\end{itemize}

Finally, an estimation of the index (\ref{ind_num}) is given by

\begin{equation}
\widehat{S}^k_{\psi} = \frac{\frac{1}{n_1}\sum_{j=1}^{n_2} \left(\psi(Y^{j};\widehat{\theta}^*) - \psi(Y^{j};\widehat{\theta}_k(X^{j}_k))\right)   }{\frac{1}{n_1}\sum_{j=1}^{n_1} \psi(Y^{j};\widehat{\theta}^*)} \label{formula_ind_num}
\end{equation}

{\rmk For higher order indices (see Remark \ref{rmk-gen_indice}) given by
\begin{equation}
S^I_{\psi} = \frac{\mathbb{E}_{(\X_I,Y)}\left(\psi(Y;\theta^*) - \psi(Y;\theta_I(\X_I))\right)}{\mathbb{E}_{Y}\psi(Y;\theta^*)} \label{ind_gen}
\end{equation}
the quantity $\theta_I(\X_{I})$ is estimated by using the sample 
$$\left(Y^{\prime\,l}_I(\X_I)=h(\X_{I^c}^{\prime\,l},\X_{I})\right)_{1\leq l \leq n_2} \, .$$}

{\rmk By considering the mean-contrast $\psi(y;\theta) = (y - \theta)^2$ we have that $\widehat{\theta}^*$ and $\widehat{\theta}_k(X^{j}_k)$ are the empirical mean of the samples $\left( Y^l = h(X^l_1,...,X^l_p)\right)_{1\leq l \leq n_1}$ and $\left(Y^{\prime\,l}_k(X^{j}_k)=h(X^{\prime\,l}_1,...,X^{\prime\,l}_{k-1},X^{j}_k,X^{\prime\,l}_{k+1},...,X^{\prime\,l}_p)\right)_{1\leq l \leq n_2}$, respectively, that is
$$ \widehat{\theta}^* = \frac{1}{n_1}\sum_{l=1}^{n_1} Y^l \,, \quad  \widehat{\theta}_k(X^{j}_k)  = \frac{1}{n_2}\sum_{l=1}^{n_2} Y^{\prime\,l}_k(X^{j}_k) \, .$$
Setting $n_1=n_2 = N$, it is easy to check that the index $\widehat{S}^k_{\psi}$ in (\ref{formula_ind_num}) is the well known Monte-Carlo estimator of the first order Sobol indice, see \cite{saltelli2002making}:
$$ \widehat{S}^k_{\psi} = \frac{\overline{\Y \Y^{\prime}_{k}} - \overline{\Y} \, \overline{\Y^{\prime}_{k}}   }{\overline{\Y^{2}} - \overline{\Y}^2}, $$
where  $\Y = (Y^1,...,Y^{N})$ and $\Y^{\prime}_{k} = (Y^{\prime\,1}_k(X^{1}_k),...,Y^{\prime\,N}_k(X^{N}_k))$ and for any vector $\uu = (u_1,...,u_N)$
$$ \overline{\uu}  = \frac{1}{N} \sum_{j=1}^{N} u_j \,.$$
}

Let us now see how  our estimation procedure works in practice.
\subsection{Numerical resolution of Example 1}

Let us remind the first example treated in Section \ref{sec-ex1} where we considered the relation
$$ Y=X_1+X_2 \, , \quad X_1\sim Exp(1), \quad  X_2\sim -X_1 $$
where $X_1$ and $X_2$ are independents. We have in Section \ref{section:laplace} analytically computed the indices $S_{\alpha}^1$ and $S_{\alpha}^2$ corresponding to the $\alpha$-quantile contrast. The obtained values are given in equations (\ref{quant1_ex1}) and (\ref{quant2_ex1}).\\
To illustrate the estimation method proposed in \ref{est}, we now compute the estimation of the indices $S_{\alpha}^1$ and $S_{\alpha}^2$ thanks to the formula (\ref{formula_ind_num}) using the  contrast $\psi_{\alpha}(y;\theta) =  (y-\theta)(\alpha -{\bf1}_{y\le\theta})$, for $\alpha$ in $[0,1]$.

\begin{figure}[htbp!]
  \centering
 \fbox{
  \includegraphics[scale=.35]{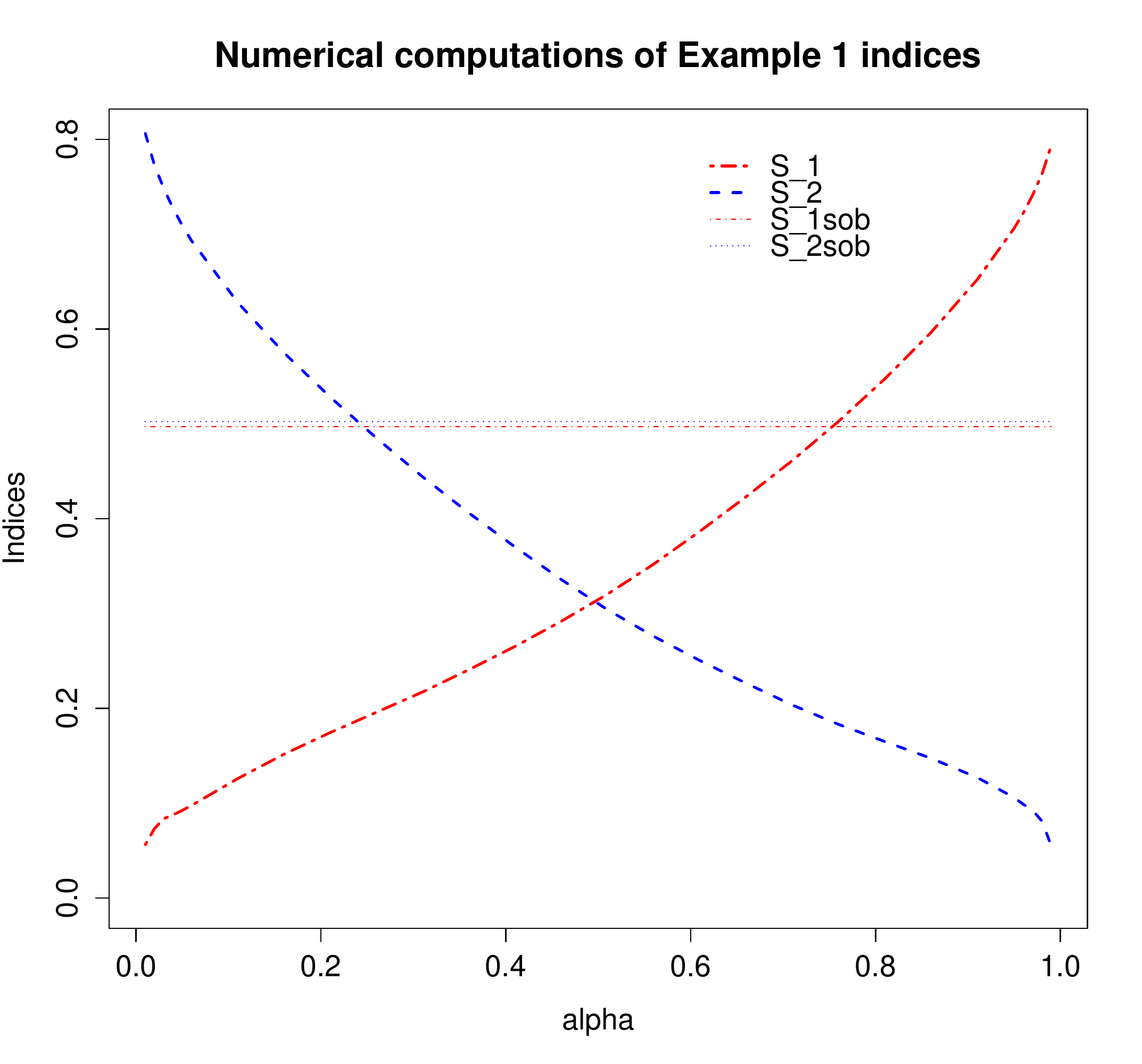}}\\
  \caption{Numerical Sensitivity Indices for Example 1.} \label{num_ex1}
\end{figure}
In Figure \ref{num_ex1} we plot the estimated values of these indices. 
Comparing Figure \ref{num_ex1} with Figure \ref{laplace}, we conclude that we obtain the same results modulo the Monte-Carlo errors due to the numerical simulation of the indices.

\subsection{Ishigami function}\label{ishi}

In this section, we consider a popular function in the domain of sensitivity analysis, the Ishigami function (see \cite{ishigami1990importance}). For instance, let us consider the following version:

$$y=\sin(\xi_1)+7\,\sin(\xi_2)^2+ 0.1\, \xi_3^4 \sin(\xi_1),$$

where the $\xi_j$'s are independent uniform random variables on $[-\pi,\pi]$. Now, we propose to analyse sensitivity indices of $\xi_1$, $\xi_2$ and $\xi_3$ by considering, first, the mean-contrast
$$ \Psi_{mean}(\theta)=\mathbb E |Y-\theta|^2 \,,$$
(we denote the corresponding indices by $S_{mean}^1,\, S_{mean}^2 $ and $S_{mean}^3$)
and then by considering the $\alpha$-quantile contrast 
$$\Psi_{\alpha}(\theta)=\mathbb E (Y-\theta)(\alpha -{\bf1}_{Y\le\theta}) \,$$
(we denote the corresponding indices by $S_{\alpha}^1,\, S_{\alpha}^2 $ and $S_{\alpha}^3$).

\begin{figure}[htbp!]
  \centering
 \fbox{
  \includegraphics[scale=.45]{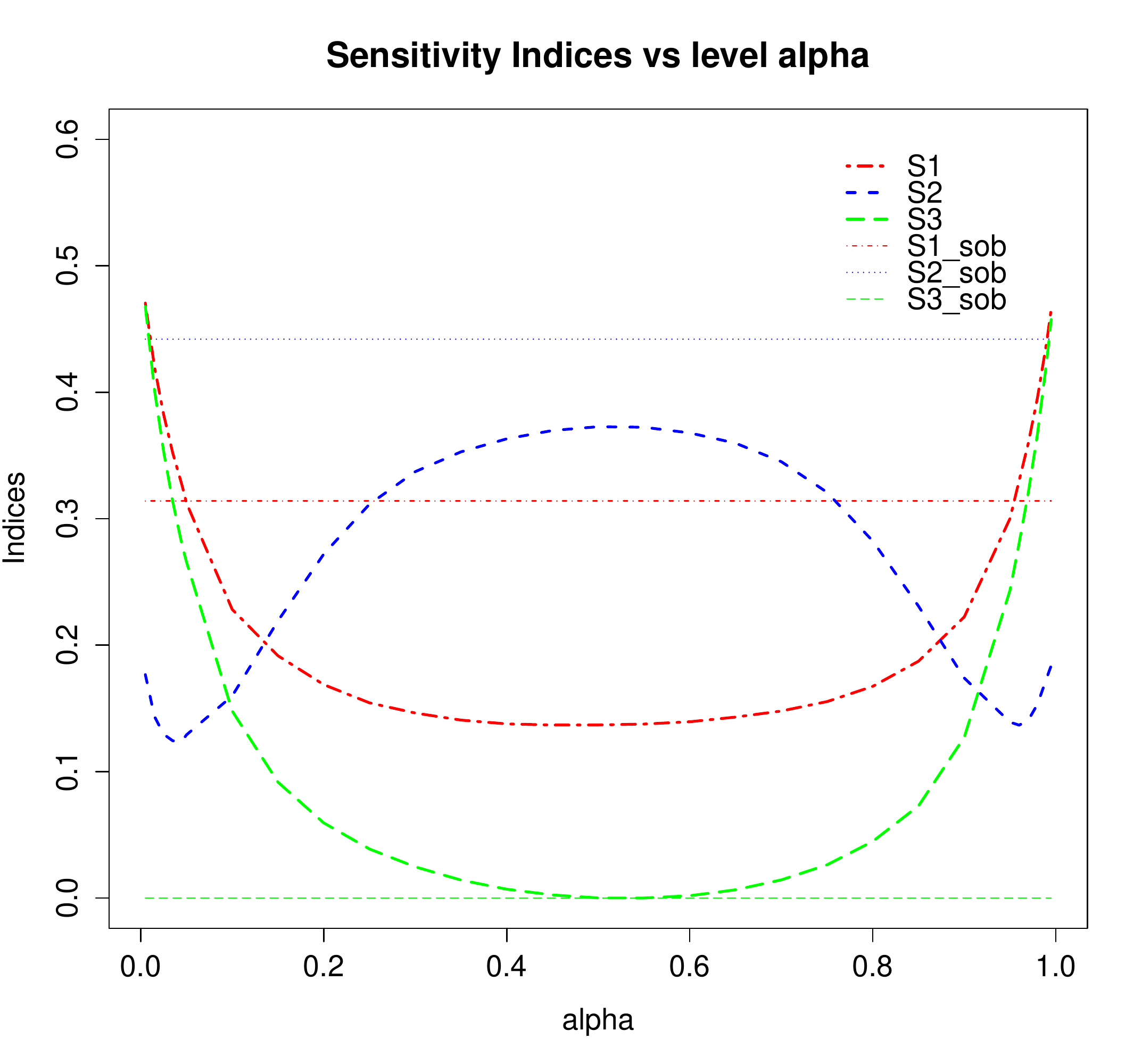}}\\
  \caption{Ishigami Sensitivity Indices}\label{figure:last}
\end{figure}

On the one hand, since the first indices $S_{mean}^1,\, S_{mean}^2 $ and $S_{mean}^3$ are the classical Sobol indices, they are known analytically and are
$$ S_{mean}^1 = 0.3139,\quad\, S_{mean}^2 = 0.4424 \quad \text{and} \quad S_{mean}^3 = 0 \,.$$
On the other hand, we do not have a closed formula of the indices $S_{\alpha}^1,\, S_{\alpha}^2 $ and $S_{\alpha}^3$, therefore we compute them numerically for $\alpha$ in $[0,1]$ and represent the results in Figure \ref{figure:last}. In this Figure one can notice the interesting point: whereas $S_{mean}^3$ is always equals to 0  which make us think that the third variable has no  influence, $S_{\alpha}^3$ is for example large for extreme values of $\alpha$  and in this case the third variable is more important than the second one\,.\\

This example shows clearly the importance of taking into account the goal of the study when measuring the impact of the input variables. Indeed when $\alpha$ is large (for example $\alpha>0.95$), one has the intuition that $\xi_3^4$ would have a significant influence. This important characteristic is detected when using the index associated with the good contrast, whereas the classical first order Sobol index is unable to do that.

\section{Conclusion}
In this paper we introduced new indices measuring the influence of an input random variable subordinated to a contrast. The range of definition of these indices is wide and allows to handle many cases that are not directly related to a variance criterion: our two main examples are the quantiles and many maximum likelihood estimators, but many others may be handled. Of course when a quadratic contrast is concerned then our indices coincide with the Sobol indices.
This paper is the first step toward a generalized theory of sensitivity analysis that embedded the Sobol index as a particular case related to variance contrasts.

\paragraph{Acknowledgements}
This work has been partially supported by the French National
Research Agency (ANR) through COSINUS program (project COSTA-BRAVA
no ANR-09-COSI-015).

\section{Annexe}
 
\subsection{Analytical computations of Example 1} \label{annexe-ex1}
Hence  the minimum value of the contrast is 
$$
\left\{\begin{matrix}
\min_\theta\Psi(\theta)&=&\Psi(\log(2\alpha))&=&\alpha(1-\log(2\alpha))& if& \alpha<1/2,\\
\min_\theta\Psi(\theta)&=&\Psi(-\log2-\log(1-\alpha))&=&(1-\alpha)\left(1-\log(2(1-\alpha))\right)& if& \alpha\geq1/2.
\end{matrix}
\right.
$$

Now we condition by $X_2$, and we compute
$$ \min_\theta\mathbb E (X_1-(\theta-X_2))(\alpha-{\bf1}_{X_1\le\theta-X_2})| X_2).$$

As the $\alpha$-quantile of a variable $Exp(1)$ is $q_{X_1}(\alpha)=-\log(1-\alpha)$, the minimum is obtained at $\theta-X_2=q_{X_1}(\alpha)$ and is $-(1-\alpha)\log(1-\alpha)$, that does not depend on $X_2$.

Following the same line for $X_1$ yields:

$$ \min_\theta\mathbb E (X_2-(\theta-X_1))(\alpha-{\bf1}_{X_2\le\theta-X_1})| X_1)=-\alpha\log(\alpha).$$
 
\subsection{Analytical computations of Example 2}\label{annexe-ex2}
 
 \begin{itemize}
\item \begin{bf}Step 1 Computation of $\min_\theta\Psi(\theta)$.\end{bf} \\\noindent
Since $\Psi(\theta)=\P(Y\geq t)-2\theta\P(Y\geq t)+\theta^2$,
the minimum value of the contrast is obtained at  $\theta=\mathbb P(Y\ge t)$. 
\begin{align*}
\mathbb P(Y\ge t)&=\begin{matrix}
\frac{ae^{-t}-e^{-at}}{a-1}\end{matrix}\\
\min_\theta\Psi(\theta)&=\P(Y\geq t)-\P(Y\geq t)^2
\end{align*}
\medskip
\item \begin{bf}Step 2 Computation of $\E\left(\min_\theta\psi(Y;\theta)\right)$.\end{bf} \\\noindent
It is obvious that $\E\left(\min_\theta\psi(Y;\theta)\right)=0$.
\medskip
\item \begin{bf}Step 3 Computation of $\E\left(\min_\theta\E\left(\psi(Y;\theta)|X_1\right)\right)$.\end{bf} \\\noindent
We have
$$
\E\left(\psi(Y;\theta)|X_1\right)=\theta^2+(1-2\theta)\E_{X_2}\left({\bf 1}_{X_2>t-X_1}\right)
$$
the minimum of this function is reached for 
$$\theta=\E_{X_2}\left({\bf 1}_{X_2>t-X_1}\right)={\bf 1}_{X_1>t}+{\bf 1}_{X_1\leq t}e^{-a(t-X_1)}.
$$
It follows that 
$$
\min_\theta\E\left(\psi(Y;\theta)|X_1\right)={\bf 1}_{X_1\leq t}\left(e^{-a(t-X_1)}-e^{-2a(t-X_1)}\right)
$$
and
$$
\E\left(\min_\theta\E\left(\psi(Y;\theta)|X_1\right)\right)=\left\{\begin{matrix}\frac{e^{-at}-e^{-t}}{1-a}-\frac{e^{-2at}-e^{-t}}{1-2a}&if &a\notin\{1,1/2\}\\
 e^{-2t}-(1-t)e^{-t}&if &a=1 \\
  2e^{-t/2}-(2+t)e^{-t}&if &a=1/2
\end{matrix}
\right.
$$
\item \begin{bf}Step 4 Computation of $\E\left(\min_\theta\E\left(\psi(Y;\theta)|X_2\right)\right)$.\end{bf} \\\noindent
We have
$$
\E\left(\psi(Y;\theta)|X_2\right)=\theta^2+(1-2\theta)\E_{X_1}\left({\bf 1}_{X_1>t-X_2}\right)
$$
the minimum of this function is reached for 
$$\theta=\E_{X_1}\left({\bf 1}_{X_1>t-X_2}\right)={\bf 1}_{X_2>t}+{\bf 1}_{X_2\leq t}e^{-(t-X_2)}.
$$
It follows that 
$$
\min_\theta\E\left(\psi(Y;\theta)|X_2\right)={\bf 1}_{X_2\leq t}\left(e^{-(t-X_2)}-e^{-2(t-X_2)}\right)
$$
and for $a\notin\{1,2\}$
$$
\E\left(\min_\theta\E\left(\psi(Y;\theta)|X_2\right)\right)=\left\{\begin{matrix}a\frac{e^{-at}-e^{-t}}{1-a}-a\frac{e^{-at}-e^{-2t}}{2-a}&if &a\notin\{1,2\}\\
e^{-2t}-(1-t)e^{-t}&if &a=1 \\
2e^{-t}- 2(t+1)e^{-2t}&if &a=2
\end{matrix}
\right.
$$

\end{itemize}

\bibliographystyle{plain}
\bibliography{sensitivity}

\end{document}